\documentclass[aps,pra,showpacs,superscriptaddress,preprint]{revtex4}
\usepackage{amssymb}
\usepackage{amsmath}
\usepackage{graphicx}

\catcode`ð=\active
 \defð{\u{g}}
 \catcode`Ð=\active
 \defÐ{\u{G}}
 \catcode`Ý=\active
\defÝ{\. I}
 \catcode`ö=\active
\defö{\"{o}}
 \catcode`Ö=\active
 \defÖ{\"O}
 \catcode`ü=\active
 \defü{\"{u}}
 \catcode`Ü=\active
 \defÜ{\"{U}}
 \catcode`Þ=\active
\defÞ{\c{S}}
 \catcode`þ=\active
 \defþ{\c{s}}
 \catcode`ý=\active
 \defý{{\i}}
 \catcode`ç=\active
\defç{\d{c}}
 \catcode`Ç=\active
\defÇ{\d{C}}

\begin{document}

\title{\textbf{Any }$l$\textbf{-state improved quasi-exact analytical
solutions of the spatially dependent mass Klein-Gordon equation for the
scalar and vector Hulth\'{e}n potentials }}
\author{Sameer M. Ikhdair}
\email[E-mail: ]{sikhdair@neu.edu.tr}
\affiliation{Department of Physics, Near East University, Nicosia, North Cyprus, Turkey}
\author{Ramazan Sever}
\email[E-mail: ]{sever@metu.edu.tr}
\affiliation{Department of Physics, Middle East Technical University, 06800, Ankara,Turkey}
\date{%
\today%
}

\begin{abstract}
We present a new approximation scheme for the centrifugal term to obtain a
quasi-exact analytical bound state solutions within the framework of the
position-dependent effective mass radial Klein-Gordon equation with the
scalar and vector Hulth\'{e}n potentials in any arbitrary $D$ dimension and
orbital angular momentum quantum numbers $l.$ The Nikiforov-Uvarov (NU)
method is used in the calculations. The relativistic real energy levels and
corresponding eigenfunctions for the bound states with different screening
parameters have been given in a closed form. It is found that the solutions
in the case of constant mass and in the case of $s$-wave ($l=0$) are
identical with the ones obtained in literature.

\end{abstract}

\maketitle

\newpage

\section{Introduction}

The bound and scattering states of the $s$- and $l$-waves for any
interaction system have raised a great interest in non-relativistic as well
as in relativistic quantum mechanics [1-3]. The exact solution of the wave
equation is very important since the wavefunction contains all the necessary
information regarding the quantum system under consideration. A number of
methods have been used to solve the wave equations exactly or quasi-exactly
for non-zero angular momentum quantum number ($l\neq 0$) by means of a given
potential. The bound state eigenvalues were solved numerically [4,5] and
quasi-analytically using variational [4,6], perturbation [7], shifted $1/N$
expansion [8,9], NU [10,11], SUSYQM [12-14] and AIM [15] methods.

The Hulth\'{e}n potential [10,12,13,15,16] is one of the important
short--range potentials in physics and it has been applied to a number of
areas such as nuclear and particle physics [17], atomic physics [18,19],
molecular physics [20,21] and chemical physics [22]. Therefore, it would be
interesting and important to solve the relativistic equation for this
potential for $l\neq 0$ case since it has been extensively used to describe
the bound and continuum states of the interaction systems. Recently, the
exact solutions for the bound and scattering states of the $s$-wave Schr\"{o}%
dinger [16,23],\ Klein-Gordon [1-3] and Dirac equation [24,25] with the
scalar and vector Hulth\'{e}n potentials are investigated.

Relativistic effects with the scalar plus vector Hulth\'{e}n-type potential
[1,2] in three- and $D$ dimensions and harmonic oscillator-type potential
[26,27] have been also discussed in the literature. The bound-states of the
Dirac and Klein-Gordon equations with the Coulomb-like scalar plus vector
potentials have been studied in arbitrary dimension [28-32]. Furthermore,
the exact results for the scattering states of the Klein-Gordon equation
with Coulomb-like scalar plus vector potentials have been investigated in an
arbitrary dimension [33].\ This equation has been exactly solved for a
larger class of linear, exponential and linear plus Coulomb potentials to
determine the bound state energy spectrum using two semiclassical methods
with the following relationship between the scalar and vector potentials: $%
V(r)=V_{0}+\beta S(r),$ $S(r)>V(r)$ where $V_{0}$ and $\beta $ being
arbitrary constants [34]. In particular, inserting the constants $V_{0}=0$
and $\beta =\pm 1$ provides the equal scalar and vector potential case $%
V(r)=\pm S(r)$.

Also, the position-dependent mass solutions of the nonrelativistic and
relativistic systems have received much attention recently. Many authors
have used different methods to study the partially exactly solvable and
exactly solvable Schr\"{o}dinger, Klein-Gordon and Dirac equations in the
presence of variable mass having a suitable mass distributions function in $%
1D,3D$ and/or any dimension $D$ cases for different potentials, such as the
exponential-type potentials [35], the Coulomb potential [36], the Lorentz
scalar interactions [37], the hyperbolic-type potentials [38], the Morse
potential [39], the P\"{o}schl-Teller potential [40], the Coulomb and
harmonic potentials [41], the modified Kratzer-type, rotationally corrected
Morse potentials [42], Mie-type and pseudoharmonic potentials [43].
Recently, the point canonical transformation (PCT) has also been employed to
solve the $D$-dimensional position-dependent effective mass Schr\"{o}dinger
equation for some molecular potentials to get the exact bound state
solutions including the energy spectrum and corresponding wave functions
[41-43].

A new method to obtain the exactly solvable PT-symmetric potential
potentials within the framework of the variable mass Dirac equation with the
vector potential coupling scheme in $(1+1)$ dimensions [38]. Three
PT-symmetric potentials are produced which are PT-symmetric harmonic
oscillator-like potential, PT-symmetric of linear plus inversely linear
potential and PT-symmetric kink-like potential. The SUSYQM formalism and
function analysis method are use to obtain the real energy levels and
corresponding spinor components for the bound states. Further, the
position-dependent effective mass Dirac equation with the PT-symmetric
hyperbolic cosecant potential can be mapped into the Schr\"{o}dinger-like
equation with the exactly solvable modified P\"{o}schl-Teller potential
[38]. The real relativistic energy levels and corresponding spinor
wavefunctions for the bound states have been given in a closed form.

The Nikiforov-Uvarov (NU) method [44] and other methods have also been used
to solve the $D$-dimensional Schr\"{o}dinger equation [45] and relativistic $%
D$-dimensional Klein-Gordon [46], Dirac [47] and spinless Salpeter equations
[48].

In strong coupling cases, it is crucial to understand relativistic effects
on a moving particle in a potential field. In a non-relativistic case, Schr%
\"{o}dinger equation with the Hulth\'{e}n potential [10,12,13,15] was solved
using the usual existing approximation, $\frac{1}{r^{2}}\approx \alpha ^{2}%
\frac{e^{\alpha r}}{\left( e^{\alpha r}-1\right) ^{2}}$ for the centrifugal
potential which was found to be consistent with the results of other methods
[4,8,13,15]. Unfortunately, this approximation is valid only for small
values of the screening parameter $\alpha ,$ but the agreement becomes poor
in the high-screening region $[10,15].$ Hence, it is of sufficient need to
improve the analytical results for the Schr\"{o}dinger equation with the
Hulth\'{e}n potential by means of a new approximation scheme. Recently,
Haouat and Chetouani [49] have solved the Klein-Gordon and Dirac equations
in the presence of the Hulth\'{e}n potential, where the energy spectrum and
the scattering wavefunctions are obtained for spin-$0$ and spin-$\frac{1}{2}$
particles, using a more general approximation scheme, $\frac{1}{r^{2}}%
\approx \alpha ^{2}\frac{e^{-\gamma \alpha r}}{\left( 1-e^{-\alpha r}\right)
^{2}}$ where $\gamma $ is a dimensionless parameter ($\gamma =0,1$ and $2$)
for the centrifugal potential. They found that the good approximation,
however, when the screening parameter $\alpha $ and the dimensionless
parameter $\gamma $ are taken as $\alpha =0.1$ and $\gamma =$1,
respectively, which is simply the case of the usual approximation
[10,12,13,15]. Also, Jia and collaborators [50] have recently proposed an
alternative approximation scheme, $\frac{1}{r^{2}}\approx \alpha ^{2}\left(
\frac{\omega }{e^{\alpha r}-1}+\frac{1}{\left( e^{\alpha r}-1\right) ^{2}}%
\right) $ where $\omega $ is a dimensionless parameter ($\omega =1.030$)$,$
for the centrifugal potential to solve the Schr\"{o}dinger equation with the
Hulth\'{e}n potential. Taking $\omega =1,$ their approximation can be
reduced into the usual approximation [10,12,13,15]. However, the accuracy of
their numerical results [50] is found to be in poor agreement with the other
numerical methods like integration and variational methods [4,5]. In order
to improve the accuracy of the used approximation, we propose and apply an
alternative shifted approximation scheme to approximate the centrifugal term
given by [51,52]

\begin{equation}
\frac{1}{r^{2}}=\underset{\alpha \rightarrow 0}{\lim }\alpha ^{2}\left[
c_{0}+\frac{e^{\alpha r}}{\left( e^{\alpha r}-1\right) ^{2}}\right] ,
\end{equation}%
where $c_{0}$ is a shifting dimensionless parameter. The approximation
scheme (1) emerged as a quite successful formalism to study the Schr\"{o}%
dinger equation with the Manning-Rosen, hyperbolic and Hulth\'{e}n
potentials in calculating the energy eigenvalues within the framework of the
NU method [51-53]. The accuracy of the results are significantly improved
over all other existing literature approximation schemes and analytical
methods [13,15,50]. With extremely high accuracy, we have obtained the
numerical energy eigenvalues as with those obtained by the numerical
integration [4,5,53], variational [4] methods and also by a MATHEMATICA
package programmed by Lucha and Sch\"{o}berl [54].

The purpose of this work is to employ the approximation scheme given in (1)
to solve the position-dependent mass radial Klein-Gordon equation\ with any
orbital angular quantum number $l$ for the scalar and vector Hulth\'{e}n
potentials in $D$-dimensions. This offers a simple, accurate and efficient
scheme for the exponential-type potential models in quantum mechanics.

Our paper is organized as follows. In section 2, we review the NU method. In
section 3, we present a brief a derivation to find the shifting parameter $%
c_{0}$. Then, the analytical solution of the position-dependent mass
Klein-Gordon equation with the scalar and vector Hulth\'{e}n potentials is
obtained for any $l$-state by means of the N-U method. Section 4 contains
the summary and conclusions.

\section{NU}

The NU method is breifly outlined here and the details can be found in [44].
This method is proposed to solve the second-order differential equation of
the hypergeometric type:
\begin{equation}
\psi _{n}^{\prime \prime }(z)+\frac{\widetilde{\tau }(z)}{\sigma (z)}\psi
_{n}^{\prime }(z)+\frac{\widetilde{\sigma }(z)}{\sigma ^{2}(z)}\psi
_{n}(z)=0,
\end{equation}%
where $\sigma (z)$ and $\widetilde{\sigma }(z)$ are polynomials, at most, of
second-degree, and $\widetilde{\tau }(s)$ is a first-degree polynomial. In
order to find a particular solution for Eq. (2), let us decompose the
wavefunction $\psi _{n}(z)$ as follows:%
\begin{equation}
\psi _{n}(z)=\phi _{n}(z)y_{n}(z).
\end{equation}%
We can reduce Eq. (2) into the form%
\begin{equation}
\sigma (z)y_{n}^{\prime \prime }(z)+\tau (z)y_{n}^{\prime }(z)+\lambda
y_{n}(z)=0,
\end{equation}%
with%
\begin{equation}
\tau (z)=\widetilde{\tau }(z)+2\pi (z),\text{ }\tau ^{\prime }(z)<0,
\end{equation}%
where $\tau ^{\prime }(z)=\frac{d\tau (z)}{dz}$ is the derivative. Also, $%
\lambda $ is a constant given in the form%
\begin{equation}
\lambda =\lambda _{n}=-n\tau ^{\prime }(z)-\frac{1}{2}n\left( n-1\right)
\sigma ^{\prime \prime }(z),\text{\ \ \ }n=0,1,2,\cdots ,
\end{equation}%
where%
\begin{equation}
\lambda =k+\pi ^{\prime }(z).
\end{equation}%
The $y_{n}(z)$ can be written in terms of the Rodrigues relation%
\begin{equation}
y_{n}(z)=\frac{B_{n}}{\rho (z)}\frac{d^{n}}{dz^{n}}\left[ \sigma ^{n}(z)\rho
(z)\right] ,
\end{equation}%
where $B_{n}$ is the normalization constant and the weight function $\rho
(z) $ satisfies%
\begin{equation}
\sigma (z)\rho ^{\prime }(z)+\left( \sigma ^{\prime }(z)-\tau (z)\right)
\rho (z)=0.
\end{equation}%
The other wavefunction in the solution is defined by%
\begin{equation}
\sigma (z)\phi ^{\prime }(z)-\pi (z)\phi (z)=0.
\end{equation}%
Further, to find the weight function in Eq. (8) we need to obtain the
following polynomial:

\begin{equation}
\pi (z)=\frac{1}{2}\left[ \sigma ^{\prime }(z)-\widetilde{\tau }(z)\right]
\pm \left\{ \frac{1}{4}\left[ \sigma ^{\prime }(z)-\widetilde{\tau }(z)%
\right] ^{2}-\widetilde{\sigma }(z)+k\sigma (z)\right\} ^{2}.
\end{equation}%
The expression under the square root sign in Eq. (11) can be arranged as the
square of a polynomial. This is possible only if its discriminant is zero.
In this regard, an equation for $k$ is being obtained. After solving such an
equation, the determined values of $k$ are included in the NU method.

\section{Bound-State Solutions}

\subsection{An Impoved Shifted Approximation Scheme}

The approximation is based on the expansion of the centrifugal term in a
series of exponentials depending on the intermolecular distance $r.$
Therefore, instead of using the usual existing approximation in literature,
let us, instead, take the following exponential-type potential to
approximate the centrifugal potential,
\begin{equation*}
\frac{1}{r^{2}}\approx \alpha ^{2}\left[ c_{0}{}+v(r)+v^{2}(r)\right] ,\text{
}v(r)=\frac{e^{\alpha r}}{e^{\alpha r}-1},
\end{equation*}%
\begin{equation}
\frac{1}{r^{2}}\approx \alpha ^{2}\left[ c_{0}+\frac{1}{e^{\alpha r}-1}+%
\frac{1}{\left( e^{\alpha r}-1\right) ^{2}}\right] .
\end{equation}%
In the low-screening region, $0.4\leq \alpha r\leq 1.2$ [15] (i.e., small
screening parameter $\alpha ),$ Eq. (12) is a very well approximation to the
centrifugal potential and the Schr\"{o}dinger equation for such an
approximation can be easily solved analytically. In Fig. 1, we give a plot
of the variation of the centrifugal potential and its approximation given in
Eq. (12) versus $\alpha r.$ It shows that the approximation (12) and $%
1/r^{2} $ are similar and coincide in both high-screening as well as in the
low-screening regions.

Changing the $r$ coordinate to $x$ by shifting the parameters as $%
x=(r-r_{0})/r_{0}$ to avoid singularities [55]$,$ we obtains%
\begin{equation}
\frac{1}{r_{0}^{2}}\left( 1+x\right) ^{-2}=\alpha ^{2}\left[ c_{0}+\frac{1}{%
e^{\gamma (1+x)}-1}+\frac{1}{\left( e^{\gamma (1+x)}-1\right) ^{2}}\right] ,%
\text{ }\gamma =\alpha r_{0},
\end{equation}%
and expanding Eq. (13) around $r=r_{0}$ $(x=0),$ we obtain the following
expansion:
\begin{equation*}
1-2x+O(x^{2})=\gamma ^{2}\left( c_{0}+\frac{1}{e^{\gamma }-1}+\frac{1}{%
\left( e^{\gamma }-1\right) ^{2}}\right)
\end{equation*}%
\begin{equation}
-\gamma ^{3}\left( \frac{1}{e^{\gamma }-1}+\frac{3}{\left( e^{\gamma
}-1\right) ^{2}}+\frac{2}{\left( e^{\gamma }-1\right) ^{3}}\right)
x+O(x^{2}),
\end{equation}%
and consequently
\begin{equation*}
\gamma ^{2}\left[ c_{0}+\frac{1}{e^{\gamma }-1}+\frac{1}{(e^{\gamma }-1)^{2}}%
\right] =1,
\end{equation*}%
\begin{equation}
\gamma ^{3}\left( \frac{1}{e^{\gamma }-1}+\frac{3}{\left( e^{\gamma
}-1\right) ^{2}}+\frac{2}{\left( e^{\gamma }-1\right) ^{3}}\right) =2.
\end{equation}%
By solving Eqs. (14) and (15) for the dimensionless parameter $c_{0},$ we
obtain
\begin{equation}
c_{0}=\frac{1}{\gamma ^{2}}-\frac{1}{e^{\gamma }-1}-\frac{1}{(e^{\gamma
}-1)^{2}}=0.0823058167837972,
\end{equation}%
where $e=2.718281828459045$ is the base of the natural logarithms and the
parameter $\gamma =0.4990429999.$

Therefore, the centrifugal potential takes the form%
\begin{equation}
\underset{\alpha \rightarrow 0}{\lim }\alpha ^{2}\left[ \frac{1}{\gamma ^{2}}%
-\frac{1}{e^{\gamma }-1}-\frac{1}{(e^{\gamma }-1)^{2}}+\frac{e^{-\alpha r}}{%
1-e^{-\alpha r}}+\left( \frac{e^{-\alpha r}}{1-e^{-\alpha r}}\right) ^{2}%
\right] =\frac{1}{r^{2}}.
\end{equation}%
Let us remark at the end of this analysis that the approximation used in
many papers in literature [10,12,13,15] is a special case of Eq. (12) if $%
c_{0}$ is set to zero.

\subsection{A Quasi-Exactly Energy Eigenvalues and Eigenfunctions}

The $D$-dimensional time-independent radial position-dependent mass
Klein-Gordon equation with scalar and vector potentials $S(r)$ and $V(r),$
respectively, $r=\left\vert \mathbf{r}\right\vert ,$ and position-dependent
mass $m(r)$ describing a spin-zero particle takes the general form [3,46]
\begin{equation*}
\mathbf{\nabla }_{D}^{2}\psi _{l_{1}\cdots l_{D-2}}^{(l_{D-1}=l)}(\mathbf{x}%
)+\frac{1}{\hbar ^{2}c^{2}}\left\{ \left[ E_{nl}-V(r)\right] ^{2}-\left[
m(r)c^{2}+S(r)\right] ^{2}\right\} \psi _{l_{1}\cdots l_{D-2}}^{(l_{D-1}=l)}(%
\mathbf{x})=0,
\end{equation*}%
\begin{equation}
\text{ }\nabla _{D}^{2}=\sum\limits_{j=1}^{D}\frac{\partial ^{2}}{\partial
x_{j}^{2}},\text{ }\psi _{l_{1}\cdots l_{D-2}}^{(l_{D-1}=l)}(\mathbf{x}%
)=R_{l}(r)Y_{l_{1}\cdots l_{D-2}}^{(l)}(\theta _{1},\theta _{2},\cdots
,\theta _{D-1}),
\end{equation}%
where $E_{nl}$ denotes the Klein-Gordon energy and $\mathbf{\nabla }_{D}^{2}$
denotes the $D$-dimensional Laplacian. Further, $\mathbf{x}$ is a $D$%
-dimensional position vector. Let us decompose the radial wavefunction $%
R_{l}(r)$ as follows:%
\begin{equation}
R_{l}(r)=r^{-(D-1)/2}g(r),
\end{equation}%
we, then, reduce Eq. (18) into the following $D$-dimensional radial
position-dependent effective mass Schr\"{o}dinger-like equation

\begin{equation}
\frac{d^{2}g(r)}{dr^{2}}+\frac{1}{\hbar ^{2}c^{2}}\left\{ \left[ E_{nl}-V(r)%
\right] ^{2}-\left[ m(r)c^{2}+S(r)\right] ^{2}-\frac{(D+2l-1)(D+2l-3)\hbar
^{2}c^{2}}{4r^{2}}\right\} g(r)=0.
\end{equation}%
Further, taking the vector and scalar potentials as the Hulth\'{e}n
potentials%
\begin{equation}
V(r)=-\frac{V_{0}e^{-\alpha r}}{1-e^{-\alpha r}},\text{ }S(r)=-\frac{%
S_{0}e^{-\alpha r}}{1-e^{-\alpha r}},\text{ }\alpha =r_{0}^{-1},\text{ }
\end{equation}%
and choosing the following mass function%
\begin{equation}
m(r)=m_{0}+\frac{m_{1}e^{-\alpha r}}{1-e^{-\alpha r}},
\end{equation}%
we can rewrite Eq. (20) as

\begin{equation*}
g^{\prime \prime }(r)+\frac{1}{\hbar ^{2}c^{2}}\left\{ \frac{2\left[
m_{0}c^{2}\left( S_{0}-m_{1}c^{2}\right) +E_{nl}V_{0}\right] e^{-\alpha r}}{%
1-e^{-\alpha r}}\right.
\end{equation*}%
\begin{equation*}
+\left. \frac{\left[ V_{0}^{2}-\left( S_{0}-m_{1}c^{2}\right) ^{2}\right]
e^{-2\alpha r}-\frac{\hbar ^{2}c^{2}\alpha ^{2}}{4}(D+2l-1)(D+2l-3)e^{-%
\alpha r}}{\left( 1-e^{-\alpha r}\right) ^{2}}\right\} g(r)
\end{equation*}%
\begin{equation}
=\frac{1}{\hbar ^{2}c^{2}}\left[ \left( m_{0}c^{2}\right)
^{2}-E_{nl}^{2}+\Delta E_{l}\right] g(r),\text{ }g(0)=0,
\end{equation}%
with the shift energy $\Delta E_{l}=\hbar ^{2}c^{2}\alpha
^{2}(D+2l-1)(D+2l-3)c_{0}/4.$ On account of the wave function $g(r)$
satisfying the standard bound-state condition (real values), i.e., $%
g(r\rightarrow \infty )\rightarrow 0.$ If we rewrite Eq. (23) by using a new
variable of the form $z=e^{-\alpha r}$ $(r\in \lbrack 0,\infty ),$ $z\in
\lbrack 1,0]),$ we get%
\begin{equation*}
\frac{d^{2}g(z)}{dz^{2}}+\frac{1-z}{z(1-z)}\frac{dg(z)}{dz}+\frac{1}{\left[
z(1-z)\right] ^{2}}
\end{equation*}%
\begin{equation}
\times \left\{ -\varepsilon _{nl}^{2}+(\beta _{1}-\beta _{4}-\gamma
+2\varepsilon _{nl}^{2})s-(\beta _{1}+\beta _{2}+\beta _{3}-\beta
_{4}+\varepsilon _{nl}^{2})s^{2}\right\} g(z)=0,
\end{equation}%
where the following definitions of parameters
\begin{equation*}
\varepsilon _{nl}=\frac{\sqrt{\left( m_{0}c^{2}\right)
^{2}-E_{nl}^{2}+\Delta E_{l}}}{Q},\text{ \ }\beta _{1}=\frac{2\left(
m_{0}c^{2}S_{0}+E_{nl}V_{0}\right) }{Q^{2}},\text{ }\beta _{2}=\frac{%
S_{0}^{2}-V_{0}^{2}}{Q^{2}},
\end{equation*}%
\begin{equation}
\beta _{3}=\frac{m_{1}c^{2}\left( m_{1}c^{2}-2S_{0}\right) }{Q^{2}},\text{ }%
\beta _{4}=\frac{2m_{0}m_{1}c^{4}}{Q^{2}},\text{ }\gamma =\frac{%
(D+2l-1)(D+2l-3)}{4},\text{ }Q=\hbar c\alpha ,
\end{equation}%
are used. For bound-state solutions, we require that $V_{0}\leq
(S_{0}-m_{1}c^{2})$ and $E_{nl}\leq \sqrt{(m_{0}c^{2})^{2}+\Delta E_{l}}.$
In order to solve Eq. (24) by means of the N-U method, we should compare it
with Eq. (2). The following values for parameters are found%
\begin{equation}
\widetilde{\tau }(z)=1-z,\text{\ }\sigma (z)=z-z^{2},\text{\ }\widetilde{%
\sigma }(z)=-\varepsilon _{nl}^{2}+(\beta _{1}-\beta _{4}-\gamma
+2\varepsilon _{nl}^{2})s-(\beta _{1}+\beta _{2}+\beta _{3}-\beta
_{4}+\varepsilon _{nl}^{2})s^{2}.
\end{equation}%
If we insert these values of parameters into Eq. (11), with $\sigma ^{\prime
}(z)=1-2z,$ the following linear function is obtained
\begin{equation}
\pi (z)=-\frac{z}{2}\pm \frac{1}{2}\sqrt{\left[ 1+4(\beta _{1}+\beta
_{2}+\beta _{3}-\beta _{4}+\varepsilon _{nl}^{2}-k)\right] z^{2}+\left[
4(k-\beta _{1}+\beta _{4}+\gamma -2\varepsilon _{nl}^{2})\right]
z+4\varepsilon _{nl}^{2}}.
\end{equation}%
The determinant of the square root must be set equal to zero, that is, $%
\Delta =(k-\beta _{1}+\beta _{4}+\gamma -2\varepsilon
_{nl}^{2})^{2}-\varepsilon _{nl}^{2}\left[ 1+4(\beta _{1}+\beta _{2}+\beta
_{3}-\beta _{4}+\varepsilon _{nl}^{2}-k)\right] =0.$ Thus, the constant $k$
found to be%
\begin{equation}
k=\beta _{1}-\beta _{4}-\gamma \pm \varepsilon _{nl}\sqrt{1+4(\beta
_{2}+\beta _{3}+\gamma )}.
\end{equation}%
In this regard, we can find four possible functions for $\pi (z)$ as
\begin{equation}
\pi (s)=-\frac{z}{2}\pm \left\{
\begin{array}{cc}
\varepsilon _{nl}\mp \left[ \varepsilon _{nl}-\frac{1}{2}\sqrt{1+4b}\right] z
& \text{\ for }k_{1}=d+\varepsilon _{nl}\sqrt{1+4b}, \\
\varepsilon _{nl}\mp \left[ \varepsilon _{nl}+\frac{1}{2}\sqrt{1+4b}\right] z
& \text{\ for }k_{2}=d-\varepsilon _{nl}\sqrt{1+4b}.%
\end{array}%
\right.
\end{equation}%
where $b=\beta _{2}+\beta _{3}+\gamma $ and $d=\beta _{1}-\beta _{4}-\gamma
. $ Thus, taking the following values%
\begin{equation}
k=\beta _{1}-\beta _{4}-\gamma -\varepsilon _{nl}\sqrt{1+4(\beta _{2}+\beta
_{3}+\gamma )},
\end{equation}%
and%
\begin{equation}
\pi (z)=-\frac{z}{2}+\varepsilon _{nl}-\left[ \varepsilon _{nl}+\frac{1}{2}%
\sqrt{1+4(\beta _{2}+\beta _{3}+\gamma )}\right] z,
\end{equation}%
they give%
\begin{equation*}
\tau (z)=1+2\varepsilon _{nl}-2\left[ 1+\varepsilon _{nl}+\frac{1}{2}\sqrt{%
1+4(\beta _{2}+\beta _{3}+\gamma )}\right] z,
\end{equation*}%
\begin{equation}
\tau ^{\prime }(s)=-2\left[ 1+\varepsilon _{nl}+\frac{1}{2}\sqrt{1+4(\beta
_{2}+\beta _{3}+\gamma )}\right] <0.
\end{equation}%
Eqs. (30)-(32) together with the assignments given in Eq. (26), the
following expressions for $\lambda $ are obtained%
\begin{equation}
\lambda _{n}=\lambda =n^{2}+\left[ 1+2\varepsilon _{nl}+\sqrt{1+4(\beta
_{2}+\beta _{3}+\gamma )}\right] n,\text{ }(n=0,1,2,\cdots ),
\end{equation}%
\begin{equation}
\lambda =\beta _{1}-\beta _{4}-\gamma -\frac{1}{2}(1+2\varepsilon _{nl})%
\left[ 1+\sqrt{1+4(\beta _{2}+\beta _{3}+\gamma )}\right] ,
\end{equation}%
where $n$ is the radial quantum number. By defining

\begin{equation}
\delta =\frac{1}{2}\left( 1+\sqrt{1+4(\beta _{2}+\beta _{3}+\gamma )}\right)
,
\end{equation}%
where $\beta _{2}+\beta _{3}=\delta ^{2}-\delta -\gamma .$ With the aid of
Eq. (35), we can easily obtain the energy eigenvalue equation of the Hulth%
\'{e}n potential by solving Eqs. (33) and (34):
\begin{equation*}
\varepsilon _{nl}^{(D)}=\frac{\left( \beta _{1}-\beta _{4}-\gamma
-n^{2}\right) -(2n+1)\delta }{2(n+\delta )}
\end{equation*}%
\begin{equation*}
=\frac{4\left[ \beta _{1}-\beta _{4}-n^{2}-(2n+1)\delta \right]
-(D+2l-1)(D+2l-3)}{8(n+\delta )}
\end{equation*}%
\begin{equation}
=\frac{2\left[ m_{0}c^{2}\widetilde{S}_{0}+E_{nl}^{\pm }V_{0}\right] +%
\widetilde{S}_{0}^{2}-V_{0}^{2}}{2Q^{2}(n+\delta )}-\frac{n+\delta }{2},%
\text{ }(n=0,1,2,\cdots ),
\end{equation}%
where $\widetilde{S}_{0}=S_{0}-m_{1}c^{2}$ is the modified scalar potential.
Solving the last equation for the energy eigenvalues $E_{nl}^{\pm },$ we
obtain

\begin{equation*}
E_{nl}^{\pm }=\frac{V_{0}}{2}\left[ 1-\frac{4\widetilde{S}_{0}\left(
\widetilde{S}_{0}+2m_{0}c^{2}\right) }{4V_{0}^{2}+\kappa _{nl}^{2}}\right]
\pm \frac{\kappa _{nl}}{2}\sqrt{\text{ }\xi -\frac{1}{4}\left[ 1-\frac{4%
\widetilde{S}_{0}\left( \widetilde{S}_{0}+2m_{0}c^{2}\right) }{%
4V_{0}^{2}+\kappa _{nl}^{2}}\right] ^{2}},
\end{equation*}%
\begin{equation*}
\xi =\frac{(2m_{0}c^{2})^{2}+\hbar ^{2}c^{2}\alpha ^{2}(D+2l-1)(D+2l-3)c_{0}%
}{4V_{0}^{2}+\kappa _{nl}^{2}},
\end{equation*}%
\begin{equation}
\kappa _{nl}=\hbar c\alpha \left( 2n+1\right) +\sqrt{4\left( \widetilde{S}%
_{0}^{2}-V_{0}^{2}\right) +\left( \hbar c\alpha \right) ^{2}\left(
D+2l-2\right) ^{2}},
\end{equation}%
where $n=0,1,2,\cdots $ and $l=0,1,2,\cdots $ signify the usual radial and
angular momentum quantum numbers, respectively, and
\begin{equation}
(\hbar c\alpha )^{2}(D+2l-2)^{2}+4\widetilde{S}_{0}^{2}\geq 4V_{0}^{2},\text{
}4\xi \geq \left[ 1-\frac{4\widetilde{S}_{0}\left( \widetilde{S}%
_{0}+2m_{0}c^{2}\right) }{4V_{0}^{2}+\kappa _{nl}^{2}}\right] ^{2},
\end{equation}%
are constraints over the strength of the potential coupling parameters. In
the above equation, let us remark that it is not difficult to conclude that
all bound-states appear in pairs, two energy solutions are valid for the
particle $E^{p}=E_{nl}^{+}$ and the second one corresponds to the
anti-particle energy $E^{a}=E_{nl}^{-}$ in the Hulth\'{e}n field. When we
take the scalar and vector potentials as $\widetilde{S}_{0}=0$ (i.e., $%
S_{0}=m_{1}c^{2}$) and $V_{0}\neq 0$, the energy equation (37) becomes
\begin{equation*}
E_{nl}^{\pm }=\frac{V_{0}}{2}\pm \frac{\kappa _{nl}}{2}\sqrt{\frac{%
(2m_{0}c^{2})^{2}+\hbar ^{2}c^{2}\alpha ^{2}(D+2l-1)(D+2l-3)c_{0}}{%
4V_{0}^{2}+\kappa _{nl}^{2}}-\frac{1}{4}},
\end{equation*}%
\begin{equation*}
(4m_{0}c^{2})^{2}+4\hbar ^{2}c^{2}\alpha ^{2}(D+2l-1)(D+2l-3)c_{0}\geq
4V_{0}^{2}+\kappa _{nl}^{2},
\end{equation*}%
\begin{equation}
\kappa _{nl}=\hbar c\alpha (2n+1)+\sqrt{\left( \hbar c\alpha \right)
^{2}\left( D+2l-2\right) ^{2}-4V_{0}^{2}},\text{ }D\geq 1,
\end{equation}%
with the following constraints on the coupling parameter of the vector
potential:
\begin{equation}
\left( \hbar c\alpha \right) ^{2}(D+2l-2)^{2}\geq 4V_{0}^{2},
\end{equation}%
must be fulfilled for real eigenvalues.

Therefore, having solved the $D$-dimensional position-dependent mass
Klein-Gordon equation for scalar and vector usual Hulth\'{e}n potentials, we
should make some useful remarks.

(i) For $s$-wave ($l=0$), the exact energy eigenvalues of the $1D$
Klein-Gordon equation becomes%
\begin{equation*}
E_{n}^{\pm }=\frac{V_{0}}{2}\left( 1-\frac{4\widetilde{S}_{0}\left(
\widetilde{S}_{0}+2m_{0}c^{2}\right) }{4V_{0}^{2}+\kappa _{n}^{2}}\right)
\pm \kappa _{n}\sqrt{\text{ }\frac{m_{0}^{2}c^{4}}{4V_{0}^{2}+\kappa _{n}^{2}%
}-\frac{1}{16}\left( 1-\frac{4\widetilde{S}_{0}\left( \widetilde{S}%
_{0}+2m_{0}c^{2}\right) }{4V_{0}^{2}+\kappa _{n}^{2}}\right) ^{2}},
\end{equation*}%
\begin{equation}
\kappa _{n}=\hbar c\alpha \left( 2n+1\right) +\sqrt{\left( \hbar c\alpha
\right) ^{2}+4(\widetilde{S}_{0}^{2}-V_{0}^{2})},
\end{equation}%
In order that at least one level might exist, it is necessary that the
inequalities%
\begin{equation}
\hbar ^{2}c^{2}\alpha ^{2}+4\widetilde{S}_{0}^{2}\geq 4V_{0}^{2},\text{ }%
\frac{16m_{0}^{2}c^{4}}{4V_{0}^{2}+\kappa _{n}^{2}}\geq \left( 1-\frac{4%
\widetilde{S}_{0}\left( \widetilde{S}_{0}+2m_{0}c^{2}\right) }{%
4V_{0}^{2}+\kappa _{n}^{2}}\right) ^{2},
\end{equation}%
are fulfilled. In the case $\widetilde{S}_{0}=0,$ $V_{0}\neq 0,$ the energy
spectrum (in units where $\hbar =c=1$):
\begin{equation}
E_{n}^{\pm }=\frac{V_{0}}{2}\pm \left[ \alpha \left( 2n+1\right) +\sqrt{%
\alpha ^{2}-4V_{0}^{2}}\right] \sqrt{\frac{m_{0}^{2}}{4V_{0}^{2}+\left[
\alpha \left( 2n+1\right) +\sqrt{\alpha ^{2}-4V_{0}^{2}}\right] ^{2}}-\frac{1%
}{16}},
\end{equation}%
with the following constraints on the potential coupling constant:%
\begin{equation}
16m_{0}^{2}\geq 4V_{0}^{2}+\left[ \sqrt{\alpha ^{2}-4V_{0}^{2}}+\left(
2n+1\right) \right] ^{2},\text{ }\alpha \geq 2V_{0},
\end{equation}%
are fulfilled for bound state solutions. We notice that the result given in
Eq. (43) is identical to Eq. (31) of Ref. [56]. As can be seen from Eq.
(43), there are only two lower-lying states $(n=0,1)$ for a Klein-Gordon
particle of rest mass $m_{0}=1$ and screening parameter $\alpha =1$ with
vector coupling strength $V_{0}\leq 1/2.$ As an example, one may calculate
the ground state energy for the vector coupling strength $V_{0}=\alpha /2$
as
\begin{equation}
E_{0}^{\pm }=\frac{V_{0}}{2}\left[ 1\pm \sqrt{\frac{2m_{0}^{2}}{V_{0}^{2}}-1}%
\right] .
\end{equation}%
Further, in the case of pure scalar potential ($V_{0}=0,S_{0}=m_{1}c^{2}),$
the energy spectrum%
\begin{equation}
E_{n}^{\pm }=\pm \sqrt{m_{0}^{2}c^{4}-\frac{\left( \hbar c\alpha \right)
^{2}\left( n+1\right) ^{2}}{4}},\text{ }4m_{0}^{2}c^{4}\geq \left( \hbar
c\alpha \right) ^{2}\left( n+1\right) ^{2}.
\end{equation}%
Since the Klein-Gordon equation is independent of the sign of $E_{n}$ for
scalar potentials, the wavefunctions become the same for both energy values.
If the range parameter $\alpha $ is chosen to be $\alpha =1/\lambda _{c},$
where $\lambda _{c}=\hbar /m_{0}c=1/m_{0}$ denotes the Compton wavelength of
the Klein-Gordon particle. It can be seen easily that while $%
S_{0}\rightarrow m_{1}c^{2}$ in ground state $(n=0),$ all energy eigenvalues
tend to the value $E_{0}\approx 0.866$ $m_{0}.$

(ii) For $D=3,$ the mixed scalar and vector Hulth\'{e}n potentials, the
energy eigenvalues for $l\neq 0$ are given by%
\begin{equation*}
E_{nl}^{\pm }=\frac{V_{0}}{2}\left( 1-\frac{4\widetilde{S}_{0}\left(
\widetilde{S}_{0}+2m_{0}c^{2}\right) }{4V_{0}^{2}+\widetilde{\kappa }%
_{nl}^{2}}\right) \pm \widetilde{\kappa }_{nl}\sqrt{\widetilde{\xi }-\frac{1%
}{4}\left( 1-\frac{4\widetilde{S}_{0}\left( \widetilde{S}_{0}+2m_{0}c^{2}%
\right) }{4V_{0}^{2}+\widetilde{\kappa }_{nl}^{2}}\right) ^{2}},
\end{equation*}%
\begin{equation*}
\widetilde{\xi }=\frac{(m_{0}c^{2})^{2}+\hbar ^{2}c^{2}\alpha ^{2}l(l+1)c_{0}%
}{4V_{0}^{2}+\widetilde{\kappa }_{nl}^{2}},
\end{equation*}%
\begin{equation}
\widetilde{\kappa }_{nl}=\hbar c\alpha \left( 2n+1\right) +\sqrt{\left(
\hbar c\alpha \right) ^{2}\left( 2l+1\right) ^{2}+4\left( \widetilde{S}%
_{0}^{2}-V_{0}^{2}\right) }.
\end{equation}%
Further, in order that at least one real eigenvalue might exist, it is
necessary that the inequality%
\begin{equation}
(\hbar c\alpha )^{2}(2l+1)^{2}+4\widetilde{S}_{0}^{2}\geq 4V_{0}^{2},\text{ }%
4\widetilde{\xi }\geq \left( 1-\frac{4\widetilde{S}_{0}\left( \widetilde{S}%
_{0}+2m_{0}c^{2}\right) }{4V_{0}^{2}+\widetilde{\kappa }_{nl}^{2}}\right)
^{2},
\end{equation}%
must be fulfilled. For the case where $\widetilde{S}_{0}=0$ in the
spatial-dependent mass ($S_{0}=0,$ in the constant mass case) [46], the
energy eigenvalues turn out to be
\begin{equation*}
E_{nl}^{\pm }=\frac{V_{0}}{2}\pm \eta _{nl}\sqrt{\frac{(m_{0}c^{2})^{2}+%
\hbar ^{2}c^{2}\alpha ^{2}l(l+1)c_{0}}{4V_{0}^{2}+\eta _{nl}^{2}}-\frac{1}{16%
}},
\end{equation*}%
\begin{equation}
\text{ }\eta _{nl}=\hbar c\alpha \left( 2n+1\right) +\sqrt{\left( \hbar
c\alpha \right) ^{2}\left( 2l+1\right) ^{2}-4V_{0}^{2}},\text{ }\hbar
c\alpha (2l+1)\geq 2V_{0},
\end{equation}%
with the following constraint over the potential parameters:%
\begin{equation}
(4m_{0}c^{2})^{2}+16\hbar ^{2}c^{2}\alpha ^{2}l(l+1)c_{0}\geq 4V_{0}^{2}+
\left[ \hbar c\alpha (2n+1)+\sqrt{\left( \hbar c\alpha \right)
^{2}(2l+1)^{2}-4V_{0}^{2}}\right] ^{2}.
\end{equation}%
(iii) When $D=3$ and $l=0,$ the centrifugal term $\frac{(D+2l-1)(D+2l-3)}{%
4r^{2}}=0$ and consequently the approximation term $\frac{%
(D+2l-1)(D+2l-3)\alpha ^{2}}{4}\left[ c_{0}+\frac{e^{-\alpha r}}{\left(
1-e^{-\alpha r}\right) ^{2}}\right] =0,$ too. Thus, the energy eigenvalues
turn to become%
\begin{equation*}
\sqrt{(m_{0}c^{2})^{2}-E_{n}^{\pm 2}}=\frac{2\left[ m_{0}c^{2}\widetilde{S}%
_{0}+E_{n}^{\pm }V_{0}\right] +\widetilde{S}_{0}^{2}-V_{0}^{2}}{2\hbar
c\alpha (n+\delta )}-\hbar c\alpha \left( \frac{n+\delta }{2}\right) ,
\end{equation*}

\begin{equation}
\delta =\frac{1}{2}\left[ 1+\frac{1}{\left( \hbar c\alpha \right) }\sqrt{%
\left( \hbar c\alpha \right) ^{2}+4\left( \widetilde{S}_{0}^{2}-V_{0}^{2}%
\right) }\right] ,\text{ (}n=0,1,2,3,\cdots )
\end{equation}%
which gives%
\begin{equation*}
E_{n}^{\pm }=\frac{V_{0}}{2}\left( 1-\frac{4\widetilde{S}_{0}\left(
\widetilde{S}_{0}+2m_{0}c^{2}\right) }{4V_{0}^{2}+\xi _{n}^{2}}\right) \pm
\varsigma _{n}\sqrt{\text{ }\frac{(m_{0}c^{2})^{2}}{4V_{0}^{2}+\varsigma
_{n}^{2}}-\frac{1}{4}\left( 1-\frac{4\widetilde{S}_{0}\left( \widetilde{S}%
_{0}+2m_{0}c^{2}\right) }{4V_{0}^{2}+\varsigma _{n}^{2}}\right) ^{2}},
\end{equation*}%
\begin{equation*}
\varsigma _{n}=\hbar c\alpha \left( 2n+1\right) +\sqrt{\left( \hbar c\alpha
\right) ^{2}+4\left( \widetilde{S}_{0}^{2}-V_{0}^{2}\right) },
\end{equation*}%
\begin{equation}
(\hbar c\alpha )^{2}+4\widetilde{S}_{0}^{2}\geq 4V_{0}^{2},\text{ }%
(4m_{0}c^{2})^{2}\geq \left( 4V_{0}^{2}+\varsigma _{n}^{2}\right) \left( 1-%
\frac{4\widetilde{S}_{0}\left( \widetilde{S}_{0}+2m_{0}c^{2}\right) }{%
4V_{0}^{2}+\varsigma _{n}^{2}}\right) ^{2}
\end{equation}%
(iv) For equal scalar and vector usual Hulth\'{e}n potential (i.e., $%
S_{0}=V_{0}$), Eq. (36) with the aid of Eq. (25) can be reduced to the
relativistic energy equation (in the conventional atomic units $\hbar =c=1$)$%
:$%
\begin{equation*}
\sqrt{m_{0}^{2}+\frac{(D+2l-1)(D+2l-3)c_{0}}{4r_{0}^{2}}-E_{R}^{2}}
\end{equation*}%
\begin{equation*}
=\frac{2r_{0}V_{0}\left[ m_{0}+E_{R}-m_{1}\right] +r_{0}(m_{1}-2m_{0})m_{1}}{%
2(n+\delta )}-\frac{n+\delta }{2r_{0}},\text{ }
\end{equation*}%
\begin{equation}
\delta =\frac{1}{2}\left[ 1+\sqrt{(D+2l-2)^{2}+\left(
2r_{0}m_{1}c^{2}\right) ^{2}-8r_{0}^{2}V_{0}m_{1}c^{2}}\right] ,\text{ (}%
n=l=0,1,2,3,\cdots ),
\end{equation}%
which is Eq. (22) of Ref. [58] if the perturbed mass $m_{1}=0$ and shifting
parameter $c_{0}=0.$

(v) We discuss non-relativistic limit of the energy equation (53). When $%
V_{0}=S_{0},$ Eq. (23) reduces to a Schr\"{o}dinger-like equation for the
potential $2V(r).$ In other words, the non-relativistic limit is the Schr%
\"{o}dinger equation for the potential $-2V_{0}e^{-r/r_{0}}/\left[
1-e^{-r/r_{0}}\right] ,$ $r_{0}=\alpha ^{-1}.$ After making the parameter
replacements $m_{0}+E_{R}\rightarrow 2m_{0}$ and $m_{0}-E_{R}\rightarrow
-E_{NR}$ in Eq. (53)[58], it reduces into the non-relativistic energy
equation of Refs. [10,12,13,15,57,59]:%
\begin{equation*}
E_{NR}=\frac{\alpha ^{2}(D+2l-1)(D+2l-3)c_{0}}{8m_{0}}-\frac{1}{8m_{0}\alpha
^{2}}\left[ \frac{(2V_{0}-m_{1})(2m_{0}-m_{1})-\alpha ^{2}\left( n+\delta
\right) ^{2}}{(n+\delta )}\right] ^{2},
\end{equation*}%
\begin{equation}
\delta =\frac{1}{2}\left[ 1+\frac{1}{\alpha }\sqrt{\alpha
^{2}(D+2l-2)^{2}+\left( 2m_{1}c^{2}\right) ^{2}-8V_{0}m_{1}c^{2}}\right] ,%
\text{ (}n=l=0,1,2,3,\cdots )
\end{equation}%
which is Eq. (23) of Ref. [57] when $c_{0}$ and $m_{1}$are set to zero. It
is noted that Eq. (54) is identical to Eq. (59) of Ref. [56] for $s$-wave in
$1D$ when the potential is $2V(r),$ when $\alpha $ becomes pure imaginary,
i.e., $\alpha \rightarrow i\alpha $ and when we set $m_{0}=1,$ $m_{1}=0$ and
$c_{0}=0.$ Equation (54) can be reduced to the constant mass ($m_{1}=0$)
case in the three-dimensional Schr\"{o}dinger equation:

\begin{equation*}
E_{NR}=\frac{\alpha ^{2}}{2m_{0}}\left\{ l(l+1)c_{0}-\left[ \frac{2V_{0}m_{0}%
}{\alpha ^{2}(n+l+1)}-\frac{n+l+1}{2}\right] ^{2}\right\} ,
\end{equation*}%
which is identical to the expressions given in Refs. [50,52] when the vector
potential is taken as $2V(r),$ $c_{0}=0$ and $\omega =1$ in Ref. [50]$.$ The
numerical approximation to the energy eigenvalues in Ref. [50] for the last
energy equation was found to be more efficient than the approximation given
in Eq. (19) of Ref. [50]. Taking $V_{0}=Z\alpha e^{2}$ as in [50], we obtain

\begin{equation*}
E_{NR}=\frac{\alpha ^{2}}{2m_{0}}\left\{ l(l+1)c_{0}-\left[ \frac{%
2m_{0}Ze^{2}}{\alpha (n+l+1)}-\frac{\left( n+l+1\right) }{2}\right]
^{2}\right\} .
\end{equation*}%
\ For the $s$-wave $(l=0),$ the above energy spectrum is identical to the
factorization method [23], SUSYQM [12,13] and NU [46] methods. Expanding the
energy equation (53) under the weak coupling conditions $\left[ (n+\delta
)/m_{0}r_{0}\right] ^{2}\ll 1$ and $\left[ V_{0}r_{0}/(n+\delta )\right]
^{2}\ll 1,$ retaining only the terms containing the power of $%
(1/m_{0}r_{0})^{2}$ and $(r_{0}V_{0})^{4},$ we obtain the relativistic
energy equation%
\begin{equation}
E_{R}\approx E_{NR}+m_{0}+2(2m_{0}-m_{1})\left( \frac{(2V_{0}-m_{1})}{%
2\alpha (n+\delta )}\right) ^{4},
\end{equation}%
which is simply Eq. (24) of Ref. [57], where $\delta $ is given in Eq. (54).
The first term is the non-relativistic energy and third term is the
relativistic approximation to energy.

Now, let us find the wave function $y_{n}(s),$ which is the polynomial
solution of hypergeometric-type equation. We multiply Eq. (4) by the weight
function $\rho (s)$ so that it can be rewritten in self-adjoint form [45,46]%
\begin{equation}
\left[ \omega (s)y_{n}^{\prime }(s)\right] ^{\prime }+\lambda \rho
(s)y_{n}(s)=0.
\end{equation}%
The weight function $\rho (s)$ that satisfies Eqs. (9) takes the following
form%
\begin{equation}
\rho (z)=z^{2\varepsilon _{nl}}(1-z)^{\beta },\text{ }\beta =2\delta -1
\end{equation}%
which gives the Rodrigues relation:%
\begin{equation*}
y_{nl}(z)=B_{nl}z^{-2\varepsilon _{nl}}(1-z)^{-\beta }\frac{d^{n}}{dz^{n}}%
\left[ z^{n+2\varepsilon _{nl}}(1-z)^{n+\beta }\right]
\end{equation*}%
\begin{equation}
=B_{nl}P_{n}^{(2\varepsilon _{nl},\beta )}(1-2z).
\end{equation}%
On the other hand, inserting the values of $\sigma (s),\pi (s)$ and $\tau
(s) $ given in Eqs. (26), (31) and (32) into Eq. (10), we get the other part
of the wave function%
\begin{equation}
\phi (s)=z^{\varepsilon _{nl}}(1-z)^{\delta }.
\end{equation}%
Hence, the wave function $g_{n}(z)=\phi _{n}(z)y_{n}(z)$ becomes
\begin{equation*}
g(z)=C_{nl}z^{\varepsilon _{nl}}(1-z)^{\delta }P_{n}^{(2\varepsilon
_{nl},\beta )}(1-2z)
\end{equation*}%
\begin{equation}
=C_{nl}z^{\varepsilon _{nl}^{(D)}}(1-z)^{\delta }P_{n}^{(2\varepsilon
_{nl}^{(D)},\beta )}(1-2z),\text{ }z\in \lbrack 1,0).
\end{equation}%
Finally, the radial wave functions of the Klein-Gordon equation are obtained
as%
\begin{equation}
R_{l}(r)=N_{nl}r^{-(D-1)/2}e^{-\varepsilon _{nl}^{(D)}\alpha r\text{ }%
}(1-e^{-\alpha r})^{\delta }P_{n}^{(2\varepsilon _{nl}^{(D)},\beta
)}(1-2e^{-\alpha r}),
\end{equation}%
with%
\begin{equation*}
\varepsilon _{nl}^{(D)}=\frac{1}{\hbar c\alpha }\sqrt{(m_{0}c^{2})^{2}+\frac{%
\hbar ^{2}c^{2}\alpha ^{2}(D+2l-1)(D+2l-3)c_{0}}{4}-E_{nl}^{2}},
\end{equation*}%
\begin{equation}
\beta =\frac{1}{\hbar c\alpha }\sqrt{4\left( \widetilde{S}%
_{0}^{2}-V_{0}^{2}\right) +\left( \hbar c\alpha \right) ^{2}\left(
D+2l-2\right) ^{2}},\text{ }\delta =\frac{1}{2}(1+\beta ),
\end{equation}%
where $E_{nl}$ is given in Eq. (37) and $N_{nl}$ is the radial normalization
factor. The Jacobi polynomials $P_{n}^{(2\varepsilon _{nl}^{(D)},\beta
)}(1-2e^{-\alpha r})$ [60] in the last result can be written in terms of the
hypergeometric function $_{2}F_{1}(-n,n+2\varepsilon _{nl}^{(D)}+\beta
+1,2\varepsilon _{nl}^{(D)};e^{-\alpha r})$ which gives the same result
obtained in Ref. [57].

(i) The exact radial wave functions for the $s$-wave Klein-Gordon equation
in $1D$ reduces to the following form (in $\hbar =c=1):$
\begin{equation*}
R_{n}(x)=C_{n}e^{-\sqrt{m_{0}^{2}-E_{n}^{2}}x\text{ }%
}(1-e^{-x/r_{0}})^{(1+a)/2}P_{n}^{(2r_{0}\sqrt{m_{0}^{2}-E_{n}^{2}}%
,a)}(1-2e^{-x/r_{0}}),
\end{equation*}%
\begin{equation}
a=\sqrt{1+4r_{0}^{2}\left( \widetilde{S}_{0}^{2}-V_{0}^{2}\right) },
\end{equation}%
where $E_{n}$ is given in Eq. (41). The last formula is identical to Eq.
(35) of Ref. [56] when the modified scalar potential, $\widetilde{S}_{0},$
is set to zero.

(ii) Choosing the atomic units $h/2\pi =\hbar =c=1,$ the exact radial wave
functions for the $s$-wave Klein-Gordon equation in $3D$ reduces to the
following form$:$%
\begin{equation*}
R_{n}(r)=N_{n}e^{-\sqrt{m_{0}^{2}+-E_{n}^{2}}r\text{ }%
}(1-e^{-r/r_{0}})^{(1+a)/2}P_{n}^{(2r_{0}\sqrt{m_{0}^{2}-E_{n}^{2}}%
,a)}(1-2e^{-r/r_{0}}),
\end{equation*}%
\begin{equation}
P_{n}^{(2r_{0}\sqrt{m_{0}^{2}-E_{n}^{2}}%
,a)}(1-2e^{-r/r_{0}})=_{2}F_{1}(-n,n+2r_{0}\sqrt{m_{0}^{2}-E_{n}^{2}}%
+a+1,2r_{0}\sqrt{m_{0}^{2}-E_{n}^{2}};e^{-\alpha r}),
\end{equation}%
where $E_{n}$ and $a$ are given in Eq. (52) and Eq. (63), respectively. The
last formula is identical to Eq. (22) of Ref. [57] when the perturbed mass $%
m_{1}$ is set to zero.

(iii) The quasi-exact radial wave functions for the $l$-wave Klein-Gordon
equation in $3D$ reduces to the following form (in $\hbar =c=1):$%
\begin{equation*}
R_{nl}(r)=N_{nl}e^{-\sqrt{m_{0}^{2}+\frac{l(l+1)c_{0}}{r_{0}^{2}}-E_{nl}^{2}}%
r\text{ }}(1-e^{-r/r_{0}})^{(1+a_{l})/2}P_{n}^{(2r_{0}\sqrt{m_{0}^{2}+\frac{%
l(l+1)c_{0}}{r_{0}^{2}}-E_{nl}^{2}},a_{l})}(1-2e^{-r/r_{0}}),
\end{equation*}%
\begin{equation*}
P_{n}^{(2r_{0}\sqrt{m_{0}^{2}+\frac{l(l+1)c_{0}}{r_{0}^{2}}-E_{nl}^{2}}%
,a_{l})}(1-2e^{-r/r_{0}})
\end{equation*}%
\begin{equation*}
=_{2}F_{1}(-n,n+2r_{0}\sqrt{m_{0}^{2}+\frac{l(l+1)c_{0}}{r_{0}^{2}}%
-E_{nl}^{2}}+a_{l}+1,2r_{0}\sqrt{m_{0}^{2}+\frac{l(l+1)c_{0}}{r_{0}^{2}}%
-E_{nl}^{2}};e^{-\alpha r}),
\end{equation*}%
\begin{equation}
a_{l}=\sqrt{\left( 2l+1\right) ^{2}+4r_{0}^{2}\left( \widetilde{S}%
_{0}^{2}-V_{0}^{2}\right) },
\end{equation}%
where $E_{nl}$ is given in Eq. (43) and $\alpha =r_{0}^{-1}$. It is
identical to Ref. [57] when $m_{1}=0$. The eigenfunctions in the constant
mass case are written as

\begin{equation*}
R_{nl}(r)=N_{nl}e^{-\sqrt{m_{0}^{2}+\frac{l(l+1)c_{0}}{r_{0}^{2}}-E_{nl}^{2}}%
r\text{ }}(1-e^{-r/r_{0}})^{(1+a_{l})/2}P_{n}^{(2r_{0}\sqrt{m_{0}^{2}+\frac{%
l(l+1)c_{0}}{r_{0}^{2}}-E_{nl}^{2}},b_{l})}(1-2e^{-r/r_{0}}),
\end{equation*}%
\begin{equation}
b_{l}=\sqrt{\left( 2l+1\right) ^{2}+4r_{0}^{2}\left(
S_{0}^{2}-V_{0}^{2}\right) }.
\end{equation}%
At the end of these calculations, the total wave functions of the
Klein-Gordon equation with position-dependent mass for the scalar and vector
Hulth\'{e}n potentials are%
\begin{equation*}
\psi _{l_{1}\cdots l_{D-2}}^{(l_{D-1}=l)}(\mathbf{x})=N_{nl}r^{-(D-1)/2}e^{-%
\varepsilon _{nl}^{(D)}\alpha r\text{ }}(1-e^{-\alpha r})^{\delta
}P_{n}^{(2\varepsilon _{nl}^{(D)},\beta )}(1-2e^{-\alpha r})
\end{equation*}%
\begin{equation*}
\frac{1}{\sqrt{2\pi }}\exp (\pm il_{1}\theta _{1})\prod\limits_{j=2}^{D-2}%
\sqrt{\frac{\left( 2l_{j}+j-1\right) n_{j}!}{2\Gamma \left(
l_{j}+l_{j-1}+j-2\right) }}\left( \sin \theta _{j}\right)
^{^{l_{j}-n_{j}}}P_{n_{j}}^{(l_{j}-n_{j}+(j-2)/2,l_{j}-n_{j}+(j-2)/2)}(\cos
\theta _{j})
\end{equation*}%
\begin{equation}
\sqrt{\frac{\left( 2n_{D-1}+2m^{\prime }+1\right) n_{D-1}!}{2\Gamma \left(
n_{D-1}+2m^{\prime }\right) }}\left( \sin \theta _{D-1}\right)
^{l_{D-2}}P_{n_{D-1}}^{(m^{\prime },m^{\prime })}(\cos \theta _{D-1}),
\end{equation}%
where $\varepsilon _{nl}^{(D)}$ and $\beta $ are given in Eq. (62) and $%
E_{nl}$ is given in Eq. (37) [46].

To check the accuracy of the resulting analytical expressions. We give a few
numerical real eigenvalues for some selected values of the mass $m_{0}$ and $%
m_{\text{1}}$and potential parameters $\ S_{0}$ and $V_{0}.$ In Tables 1 and
2, taking $\alpha =1$ and $m_{0}=1,$ we present some numerical values for
the energy spectrum of the constant mass Klein-Gordon equations with the
condition $S_{0}=V_{0}$ for all possible real eigenvalues. To get more real
energy eigenvalues in the constant mass case (e.g., $m_{0}=1,$ $m_{1}=0$),
the vector parameter $V_{0}$ of the Hulth\'{e}n potential should be
increased. As shown in Tables 1 and 2, when the parameter $%
V_{0}=S_{0}=1,2,3,6,8,20,$ we obtain $N=$ 1,3,6,10,15,36 real energy
eigenvalues, respectively. The numerical solution of the position-dependent
mass case with vector and scalar Hulth\'{e}n potential parameters satisfying
the conditions $S_{0}=\pm V_{0}$ and $S_{0}>V_{0}$ are presented in Table 3.
For example, in Table 3, when the Hulth\'{e}n potential parameter $%
V_{0}=S_{0}=1,$ $m_{0}=5$ and $m_{1}\neq 0,$ we obtain $N=$ 46 real energy
eigenvalues. Obviously, the number of real eigenvalues increases in the
solution of the position-dependent case than in the constant mass case where
the condition $S_{0}\geq V_{0}$ must be fulfilled.

\section{Coclusions}

In summary, we have proposed an alternative approximation scheme for the
centrifugal potential similar to the non-relativistic case. This is because
the usual approximation [10,13,15] for the centrifugal term is only valid
for low-screening region, however, for the high screening region where $%
\alpha $ increases, the agreement between the old approximation and
centrifugal term decreases. Using this approximation scheme, the analytical
solutions of the radial Klein-Gordon equation with position-dependent mass
for scalar and vector Hulth\'{e}n potentials can be approximately obtained
for any dimension $D$ and orbital angular momentum quantum number $l$. It is
found that the expressions for the eigenvalues and the corresponding
eigenfunctions become complicated and tedious since the eigenvalues are
related to the parameters $m_{o},m_{1},S_{0},V_{0},c_{0}$ and $\alpha .$ We
have investigated the possibility to obtain the bound-state (real) energy
spectra with some constraints to be imposed on the parameters and, further,
the relationship between the strengths of vector $V_{0}$ and scalar $S_{0}$
coupling parameters. In one- and three-dimensions, the special cases for the
angular momentum $l=0,1$ are carried out in detail. We find that the
analytical expressions of the energy eigenvalues and eigenfunctions are
identical with the results obtained by other methods. The analytical energy
equation and the unnormalized radial wavefunctions are expressed in terms of
hypergeometric polynomials. For constant mass case $(m_{1}=0)$ and $s$-wave (%
$l=0),$ the results are reduced to exact solution of bound states of $s$%
-wave Klein-Gordon equation with scalar and vector Hulth\'{e}n potentials.
To test our results, we have also calculated the energy eigenvalues of a
particle and antiparticle for the constant mass limit as well as the
position-dependent mass case. The case of spatial-dependent mass with scalar
potential $S_{0}=m_{1}c^{2}$ is found to be equivalent to the constant mass
with scalar potential $S_{0}=0$ in a pure vector case. Hence, the spectrum
is found to be same.

\acknowledgments Work partially supported by the Scientific and
Technological Research Council of Turkey (T\"{U}B\.{I}TAK). We thank the
kind referees for their positive and invaluable suggestions which have
improved the paper greatly.

\newpage

{\normalsize 
}

\newpage

\begin{table}[tbp]
\caption{The energy spectrum of the scalar and vector Hulth\'{e}n potential
for $m_{0}=1$ and $m_{1}=0.$}%
\begin{tabular}{lllllll}
$V_{0}=S_{0}$ & $n$ & $l$ & $E_{nl}^{+}$\tablenotemark[1]%
\tablenotetext[1]{The present NU method.} & $E_{nl}^{-}$\tablenotemark[1] & $%
E_{nl}^{+}$ [61,62]\tablenotemark[2]\tablenotetext[2]{The results from AIM
and SUSY.} & $E_{nl}^{-}$ [61,62]\tablenotemark[2] \\
\tableline$1$ & $1$ & $0$ & $1.000000$ & $-0.600000$ & $1.000000$ & $%
-0.600000$ \\
& $1$ & $1$ & $-$ & $-$ & $-$ & $-$ \\
$2$ & $1$ & $0$ & $0.707107$ & $-0.707107$ & $0.707107$ & $-0.707107$ \\
& $1$ & $1$ & $0.984171$ & $-0.214941$ & $-$ & $-$ \\
& $1$ & $2$ & $-$ & $-$ & $-$ & $-$ \\
& $2$ & $0$ & $0.984171$ & $-0.214941$ & $0.984171$ & $-0.214941$ \\
& $2$ & $1$ & $-$ & $-$ & $-$ & $-$ \\
$3$ & $1$ & $0$ & $0.302169$ & $-0.763708$ & $0.302169$ & $-0.763708$ \\
& $1$ & $1$ & $0.911438$ & $-0.411438$ & $-$ & $-$ \\
& $1$ & $2$ & $0.600000$ & $0.600000$ & $-$ & $-$ \\
& $1$ & $3$ & $-$ & $-$ & $-$ & $-$ \\
& $2$ & $0$ & $0.911438$ & $-0.411438$ & $0.911438$ & $-0.411438$ \\
& $2$ & $1$ & $0.600000$ & $0.600000$ & $-$ & $-$ \\
& $2$ & $2$ & $-$ & $-$ & $-$ & $-$ \\
& $3$ & $0$ & $0.600000$ & $0.600000$ & $0.600000$ & $0.600000$ \\
& $3$ & $1$ & $-$ & $-$ & $-$ & $-$ \\
$6$ & $1$ & $0$ & $-0.355051$ & $-0.844949$ & $-0.355051$ & $-0.844949$ \\
& $1$ & $1$ & $0.235890$ & $-0.635890$ & $-$ & $-$ \\
& $1$ & $2$ & $0.763708$ & $-0.302169$ & $-$ & $-$ \\
& $1$ & $3$ & $0.994273$ & $0.284416$ & $-$ & $-$ \\
& $2$ & $0$ & $0.235890$ & $-0.635890$ & $0.235890$ & $-0.635890$ \\
& $2$ & $1$ & $0.763708$ & $-0.302169$ & $-$ & $-$ \\
& $2$ & $2$ & $0.994273$ & $-0.284416$ & $-$ & $-$ \\
& $2$ & $3$ & $-$ & $-$ & $-$ & $-$ \\
& $3$ & $0$ & $0.763708$ & $-0.302169$ & $0.763708$ & $-0.302169$ \\
& $3$ & $1$ & $0.994273$ & $0.284416$ & $-$ & $-$ \\
& $3$ & $2$ & $-$ & $-$ & $-$ & $-$ \\
& $4$ & $0$ & $0.994273$ & $0.284416$ & $0.994273$ & $0.284416$%
\end{tabular}%
\end{table}

\bigskip

\newpage
\begin{table}[tbp]
\caption{The energy spectrum of the scalar and vector Hulth\'{e}n potential
for $m_{0}=1$ and $m_{1}=0.$}%
\begin{tabular}{llllllllll}
$V_{0}=S_{0}$ & $n$ & $l$ & $E_{nl}^{+}$ & $E_{nl}^{-}$ & $V_{0}=S_{0}$ & $n$
& $l$ & $E_{nl}^{+}$ & $E_{nl}^{-}$ \\
\tableline$8$ & $1$ & $0$ & $-0.539504$ & $-0.872260$ & $20$ & $2$ & $0$ & $%
-0.662662$ & $-0.853230$ \\
& $1$ & $1$ & $-0.063251$ & $-0.703872$ &  & $2$ & $1$ & $-0.418342$ & $%
-0.735504$ \\
& $1$ & $2$ & $0.447214$ & $-0.447214$ &  & $2$ & $2$ & $-0.127025$ & $%
-0.578857$ \\
& $1$ & $3$ & $0.870312$ & $-0.061324$ &  & $2$ & $3$ & $0.194284$ & $%
-0.377770$ \\
& $1$ & $4$ & $0.800000$ & $0.8000000$ &  & $2$ & $4$ & $0.523260$ & $%
-0.122370$ \\
& $1$ & $5$ & $-$ & $-$ &  & $2$ & $5$ & $0.825665$ & $0.208818$ \\
& $2$ & $0$ & $-0.063251$ & $-0.703872$ &  & $2$ & $6$ & $0.998229$ & $%
0.706553$ \\
& $2$ & $1$ & $0.447214$ & $-0.447214$ &  & $3$ & $0$ & $-0.418342$ & $%
-0.735504$ \\
& $2$ & $2$ & $0.870312$ & $-0.061324$ &  & $3$ & $1$ & $-0.127025$ & $%
-0.578857$ \\
& $2$ & $3$ & $0.800000$ & $0.800000$ &  & $3$ & $2$ & $0.194284$ & $%
-0.377770$ \\
& $2$ & $4$ & $-$ & $-$ &  & $3$ & $3$ & $0.523260$ & $-0.122370$ \\
& $3$ & $0$ & $0.447214$ & $-0.447214$ &  & $3$ & $4$ & $0.825665$ & $%
0.208818$ \\
& $3$ & $1$ & $0.870312$ & $-0.061324$ &  & $3$ & $5$ & $0.998229$ & $%
0.706553$ \\
& $3$ & $2$ & $0.800000$ & $0.800000$ &  & $4$ & $0$ & $-0.127025$ & $%
-0.578857$ \\
& $3$ & $3$ & $-$ & $-$ &  & $4$ & $1$ & $0.194284$ & $-0.377770$ \\
& $4$ & $0$ & $0.870312$ & $-0.061324$ &  & $4$ & $2$ & $0.523260$ & $%
-0.122370$ \\
& $4$ & $1$ & $0.800000$ & $0.800000$ &  & $4$ & $3$ & $0.825665$ & $%
0.208818 $ \\
& $4$ & $2$ & $-$ & $-$ &  & $4$ & $4$ & $0.998229$ & $0.706553$ \\
& $5$ & $0$ & $0.800000$ & $0.800000$ &  & $5$ & $0$ & $0.194284$ & $%
-0.377770$ \\
& $5$ & $1$ & $-$ & $-$ &  & $5$ & $1$ & $0.523260$ & $-0.122370$ \\
& $6$ & $0$ & $-$ & $-$ &  & $5$ & $2$ & $0.825665$ & $0.208818$ \\
$20$ & $1$ & $0$ & $-0.846811$ & $-0.935368$ &  & $5$ & $3$ & $0.998229$ & $%
0.706553$ \\
& $1$ & $1$ & $-0.662662$ & $-0.853230$ &  & $6$ & $0$ & $0.523260$ & $%
-0.122370$ \\
& $1$ & $2$ & $-0.418342$ & $-0.735504$ &  & $6$ & $1$ & $0.825665$ & $%
0.208818$ \\
& $1$ & $3$ & $-0.127025$ & $-0.578857$ &  & $6$ & $2$ & $0.998229$ & $%
0.706553$ \\
& $1$ & $4$ & $0.194284$ & $-0.377770$ &  & $7$ & $0$ & $0.825665$ & $%
0.208818$ \\
& $1$ & $5$ & $0.523260$ & $-0.122370$ &  & $7$ & $1$ & $0.998229$ & $%
0.706553$ \\
& $1$ & $6$ & $0.825665$ & $0.208818$ &  & $8$ & $0$ & $0.998229$ & $%
0.706553 $ \\
& $1$ & $7$ & $0.998229$ & $0.706553$ &  & $9$ & $0$ & $-$ & $-$%
\end{tabular}%
\end{table}

\bigskip

\begin{table}[tbp]
\caption{The energy spectrum of the scalar and vector Hulth\'{e}n potential
for $m_{1}\neq 0.$}%
\begin{tabular}{llllllllllllllll}
$m_{0}$ & $m_{1}$ & $V_{0}$ & $S_{0}$ & $n$ & $l$ & $E^{+}$ & $E^{-}$ & $%
m_{0}$ & $m_{1}$ & $V_{0}$ & $S_{0}$ & $n$ & $l$ & $E^{+}$ & $E^{-}$ \\
\tableline$5$ & $0.01$ & $2$ & $2$ & $1$ & $0$ & $0.822925$ & $-4.913410$ & $%
5$ & $1$ & $-10$ & $20$ & $1$ & $0$ & $4.857570$ & $-1.483450$ \\
&  &  &  & $1$ & $1$ & $3.110670$ & $-4.804170$ &  &  &  &  & $1$ & $1$ & $%
4.875450$ & $-1.571890$ \\
&  &  &  & $2$ & $0$ & $3.065630$ & $-4.807820$ &  &  &  &  & $2$ & $0$ & $%
4.999480$ & $-2.709050$ \\
&  &  &  & $2$ & $1$ & $4.252020$ & $-4.650830$ &  &  &  &  & $2$ & $1$ & $%
4.999990$ & $-2.772530$ \\
&  &  &  & $2$ & $2$ & $4.795730$ & $-4.445800$ &  &  &  &  & $2$ & $2$ & $%
4.998750$ & $-2.895220$ \\
&  &  &  & $3$ & $0$ & $4.229630$ & $-4.655840$ &  &  &  &  & $3$ & $0$ & $%
4.924130$ & $-3.601650$ \\
&  &  &  & $3$ & $1$ & $4.793910$ & $-4.447040$ &  &  &  &  & $3$ & $1$ & $%
4.914310$ & $-3.648140$ \\
&  &  &  & $3$ & $2$ & $4.989330$ & $-4.185200$ &  &  &  &  & $3$ & $2$ & $%
4.893220$ & $-3.737900$ \\
&  &  &  & $3$ & $3$ & $4.956220$ & $-3.857960$ &  &  &  &  & $3$ & $3$ & $%
4.858140$ & $-3.864780$ \\
$5$ & $0.01$ & $-2$ & $2$ & $1$ & $0$ & $4.913410$ & $-0.822930$ & 5 & $0.1$
& $1$ & $1$ & $1$ & $0$ & 3.443410 & -4.868720 \\
&  &  &  & $1$ & $1$ & $4.804170$ & $-3.110670$ &  &  &  &  & $1$ & $1$ & $%
4.722690$ & $-4.742880$ \\
&  &  &  & $2$ & $0$ & $4.807820$ & $-3.065630$ &  &  &  &  & $2$ & $0$ & $%
4.618770$ & $-4.768190$ \\
&  &  &  & $2$ & $1$ & $4.650830$ & $-4.252020$ &  &  &  &  & $2$ & $1$ & $%
4.982510$ & $-4.577550$ \\
&  &  &  & $2$ & $2$ & $4.445800$ & $-4.795730$ &  &  &  &  & $2$ & $2$ & $%
4.964780$ & $-4.347700$ \\
&  &  &  & $3$ & $0$ & $4.655840$ & $-4.229630$ &  &  &  &  & $3$ & $0$ & $%
4.960360$ & $-4.613290$ \\
&  &  &  & $3$ & $1$ & $4.447040$ & $-4.793910$ &  &  &  &  & $3$ & $1$ & $%
4.967570$ & $-4.354450$ \\
&  &  &  & $3$ & $2$ & $4.185200$ & $-4.989330$ &  &  &  &  & $3$ & $2$ & $%
4.788530$ & $-4.056980$ \\
&  &  &  & $3$ & $3$ & $3.857960$ & $-4.956220$ &  &  &  &  & $3$ & $3$ & $%
4.484330$ & $-3.682040$ \\
$5$ & $0.1$ & $-2$ & $5$ & $1$ & $0$ & $4.871650$ & $-3.222360$ &  &  &  &
& $4$ & $0$ & $4.984480$ & $-4.401670$ \\
&  &  &  & $1$ & $1$ & $4.926240$ & $-3.503700$ &  &  &  &  & $4$ & $1$ & $%
4.794830$ & $-4.065620$ \\
&  &  &  & $2$ & $0$ & $5.000000$ & $-4.245710$ &  &  &  &  & $4$ & $2$ & $%
4.488330$ & $-3.686650$ \\
&  &  &  & $2$ & $1$ & $4.995470$ & $-4.392630$ &  &  &  &  & $4$ & $3$ & $%
4.054980$ & $-3.206920$ \\
&  &  &  & $2$ & $2$ & $4.965180$ & $-4.615030$ &  &  &  &  & $4$ & $4$ & $%
3.455290$ & $-2.575480$ \\
&  &  &  & $3$ & $0$ & $4.915250$ & $-4.768460$ &  &  &  &  & $5$ & $0$ & $%
4.837690$ & $-4.126180$ \\
&  &  &  & $3$ & $1$ & $4.878060$ & $-4.836860$ &  &  &  &  & $5$ & $1$ & $%
4.497830$ & $-3.697630$ \\
&  &  &  & $3$ & $2$ & $4.793250$ & $-4.930300$ &  &  &  &  & $5$ & $2$ & $%
4.060510$ & $-3.212870$ \\
&  &  &  & $3$ & $3$ & $4.647670$ & $-4.993400$ &  &  &  &  & $5$ & $3$ & $%
3.459590$ & $-2.579950$ \\
&  &  &  &  &  &  &  &  &  &  &  & $5$ & $4$ & $2.567010$ & $-1.664550$ \\
&  &  &  &  &  &  &  &  &  &  &  & $5$ & $5$ & $-$ & $-$%
\end{tabular}%
\end{table}

\newpage

\begin{figure}[htbp]
\centering
\includegraphics[height=3in, width=5in, angle=0]{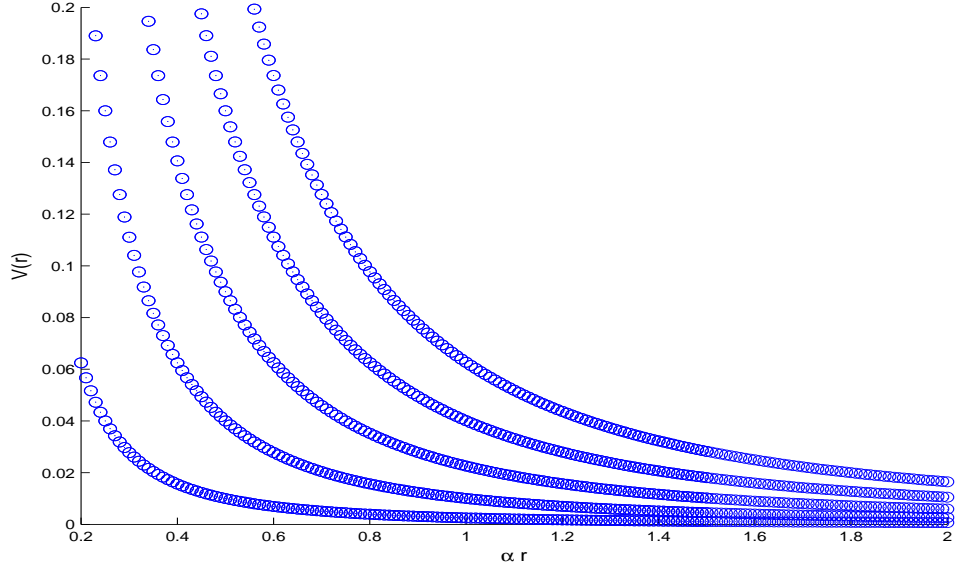}
\caption{A plot of the variation of the centrifugal potential,
$1/r^{2}$ and its corresponding propose approximation
form $\protect\alpha ^{2}\left[ c_{0}+\frac{e^{\protect\alpha r}}{\left( e^{%
\protect\alpha r}-1\right) ^{2}}\right] $\ versus $\protect\alpha
r,$ where
the screening parameter $\protect\alpha $ changes from $\protect\alpha %
=0.050 $ to $\protect\alpha =0.250$ in steps of $0.050.$}
\end{figure}

\end{document}